\documentstyle[12pt,aaspp4]{article}
 
\lefthead{Hutchings et al.}
\righthead{QSO 1345+584}
 
\begin{document}
\title{HST resolved image and spectra of z$\simeq$2 QSO 1345+584}
 
\author{J.B. Hutchings\altaffilmark{1}} 
\affil{Dominion Astrophysical Observatory\\Herzberg Institute of Astrophysics,
National Research Council of 
Canada\\ 5071 W. Saanich Rd., Victoria, B.C. V8X 4M6, Canada} 
 
\authoremail{john.hutchings@hia.nrc.ca}

\altaffiltext{1}{Based on observations with the NASA/ESA \it Hubble Space 
Telescope,
\rm obtained at the Space Telescope Science Institute, which is operated
by AURA Inc under NASA contract NAS5-26555}

\begin{abstract}

  The QSO 1345+584 has been spatially resolved by direct images and in spectral
images, and has extended flux asymmetrically to the W, where its inner radio
structure is seen. The brightest knots in the resolved flux correspond closely with
knots in the curved radio jet, and the brightest knot has velocity of approach of
some 3000 km/s with respect to the nucleus. Other parts of the line-emitting material
appear to follow a systematic velocity field with values up to 1000 km/s with 
respect to the nucleus. The signal from the resolved continuum is not detected
spectroscopically but accounts for 2/3 of the (rest UV) flux, so that it is likely 
to originate in hot stars. The QSO lies in or behind a compact group of galaxies 
of comparable brightness and irregular and knotty morphology, which probably form 
a dense physical group with very young stellar populations.  

\end{abstract}

 
\section{Introduction and data}

   The Space Telescope Imaging Spectrograph (STIS) instrument on the 
Hubble Space Telescope  (HST) offers new opportunities
for QSO host galaxy studies. The instrument may be used to obtain direct
images with the CCD which are deeper and less noisy than WFCP2, and UV
images with the MAMA detectors. Use of a long slit with STIS enables
spatially resolved spectra of the host galaxy, both through and away from 
the nucleus, with a pixel spacing of $\sim$ 0.05" in the visible and 0.025"
in the UV.

   This paper presents the results of CCD imaging and spectroscopy of the
QSO 1345+584, which has redshift 2.039. The high redshift moves rest-UV
into the visible, so little observed UV flux is expected and no MAMA
observations were attempted. The QSO was chosen for several reasons:
it has detected extended Ly$\alpha$ flux from ground-based observations (Heckman
et al 1991), with $\sim$1.5" resolution; it is a radio-loud object with
complex compact structure (Lonsdale et al 1993) that might be seen in the 
visible; and it lies
in the continuous viewing zone (CVZ) of HST so that long integrations could
be performed. 

 The QSO has a catalogued visible magnitude of 17.5 and is also known as 
4C 58.27 or OP 577 (Hewitt and Burbidge 1993). 
Lehnert et al (1993) were unable to detect any resolved continuum flux
around the QSO, in ground-based observations. Barthel et al (1988) noted
that the bend angle of the structure is remarkable and suggests stong interaction
with the surrounding medium. Bremer et al (1992) describe long-slit spectra
in which no definite extended line emission was detected.    
  
The observations were performed on 
1997 Dec 19 as shown in Table 1. The short image exposure was taken to
obtain the nucleus without detector saturation. The roll angle was 
determined by the CVZ window, and spectra were taken centred on the nucleus
(slit A) and also offset arbitrarily to the north by 1.5" (slit B), through 
a 2" wide slit. The wide slit degrades the 
spectral resolution but was judged necessary to optimise detection of the
continuum signal from the faint QSO host galaxy. All observations 
were performed with 3 CCD readouts, to identify cosmic rays. 
Standard wavelength calibration exposures were taken with the spectra.
 
  All the images were processed with standard CALSTIS reduction, but
using a mean dark image from the day of observation to eliminate hot pixels
as well as possible, since the signal levels of interest are low. The
individual reads were combined to eliminate cosmic rays in the standard way.
We discuss below the results and measurements made in the processed data.

\section{Image}

   The direct image  of the QSO and immediate surroundings is shown in
Figure~\ref{fig1}. The QSO is the brightest object in the field and shows 
diffraction spikes and a reflection ring to the E which is caused by a 
reflection off the detector window. The STIS read noise, wide bandpass, and 
DQE make this image comparable in depth with a WFPC2 broad-band image of
some 10 times the exposure, or 20000 sec. Point sources to 30 mag are
detected with 3$\sigma$ confidence, as surface brightnesses to $\sim$27
mag/arcsec$^2$. While the surface brightness limit is not deep compared
with large ground-based telescopes, it applies over areas significantly
smaller than one arcsec.

   The ring to the E of the QSO that looks like a curved arm is the
internal camera reflection, as are some small radial features around the
point. The exact structure of the PSF, particularly the reflection ring, 
depends on the location within the image. Suitably placed stellar images were
found in the database of parallel images for use in removing the PSF
from the QSO image. Figure~\ref{fig2} shows the PSF-removed image. The signal level
in the ring is some 10 times higher than the faintest structure detectable
in the clean parts of the image, and the noise and uncertainty of the
subtraction means that we know little of the real structure in this region.
However, there is definite detected structure at distances several times 
larger than the ring, mostly on the opposite (W) side. This agrees with the
ground-based result, as shown in Figure~\ref{fig3}. The ground-based result goes to
fainter fluxes, but lacks information in the inner part and has some 30
times lower spatial resolution.

   Comparison of the visible structure with the radio shows a high degree
of coincidence, if we align the eastern radio knot with the optical nucleus,
as also shown in Figure~\ref{fig3}. 
The correspondence is not exact in position or relative brightness. 
With the exception of the inner bright knot, the radio knots are more compact
than the optical. The optical flux at the western-most radio knot is weaker and
lies outside the radio: this is where it curves most strongly. The outer
two knots to the north lie very close to the optical. In addition to the
radio knots, there are other regions of optical flux. 

   The QSO nucleus has m$_V\sim$ 18.5, which is fainter than the 17.5 quoted in
Hewitt and Burbidge (1993), but consistent with the m$_B$=18.6 quoted by
Heckman et al (1991), and the very blue continuum seen in our spectrum. 
The resolved flux from the host galaxy may be measured in 
two ways. First, the azimuthally averaged luminosity profile is compared with that
of the PSF normalised to the peak value, and the flux of the difference measured.
Second the flux from the continuum-subtracted image is measured (Fig 2).
In the first instance, the central pixels of the QSO are saturated in the
long exposure, and were replaced by scaled values of the short exposure.
The difference flux  down to a level of 26 mag/arcsec$^2$ is 31 times less
than the PSF, or magnitude 22.4. The flux from the PSF-subtracted QSO is
less by 25\%. This reaches to similar flux levels, but is subject to innacuracies 
near the bright core, and this region was omitted from the measurement. 
Thus a good estimate of the host magnitude is 22.4, which corresponds to absolute
magnitude $\sim$-22 if it is a star-forming galaxy (assuming H$_0$=65 and q$_0$=0.5).
However, we show below that
much of the blue flux is Ly$\alpha$ line emission, and since C IV is the only other
strong line feature in the CCD bandpass (rest wavelength $\sim$1200 to 2900A),
the galaxy continuum may have M$_V$ nearer -21.5.

   A luminosity plot (Figure~\ref{fig4})  
of a smoothed (gaussian FWHM 2.5 pixels) image shows
the QSO profile continues to fall over 80 pixels (4 arcsec) which is comparable
with the ground-based detection contours (Heckman et al 1991). A contour plot
of the smoothed image (Figure 3) shows the faint flux is not symmetrical, but more
extended to the W, as also found in the ground-based data. Similar faint haloes
appear around the other bright galaxies, so that their outer limits fill a
good fraction ($\sim$25\%) of the sky in this group.

   The image also shows a group of remarkable looking galaxies in the field.
They are not evenly distributed, lying almost entirely to the W of the
QSO. Table 2 shows the photometric measures of the principal galaxies, as
labelled in Figure~\ref{fig1}. The bright objects have similar sizes and total
flux, which are similar to the QSO host galaxy. This kind of group is unusual: 
in comparing randomly selected images from the parallel exposure program, of 7 
other high latitude fields with STIS, 
with similar exposures, we find no other case of a concentration within the field,
of galaxies with this small brightness range. (The closest
cases are a group of smooth galaxies associated with a large bright galaxy, 
presumably at low redshift, and in a deep
image (2100 sec) a few non-clustered compact knotty galaxies, presumably at high z.
The other random images are devoid of any galaxies of this brightness.)
In the 1345+584 field, the galaxy morphologies are varied, but mostly
irregular and knotty - very like rest UV images of galaxies with active
star-formation. Galaxy 5 is a face-on spiral with a faint nucleus and
bulge but with very bright knots in the spiral arms. Galaxy 1 is edge-on
with no bulge or strong dust lane, and a knot in the plane SW of the nucleus.
Galaxy 2 lies next to a star, but has irregular morphology and several
bright knots, none of which is central. Galaxy 6 has an elliptical shape but 
contains 3 bright knots as well as a resolved nuclear region. Galaxy 4 is compact
and bright, irregular, and appears to be associated with a curved string
of knots to the NE, the inner ones of which connect to the main galaxy.
There are numerous other galaxies with a range of size and surface brightness,
and almost all irregular or knotty.

Thus, it seems possible that these galaxies are companions of the QSO at z=2.0, seen 
in rest UV wavelengths. Sizes at this redshift are in the range 5 to 20 Kpc,
and the entire group lies within $\sim$150 Kpc projected on the sky. 
If they are at the QSO redshift and young, k-corrections will be negative 
(-1.5 for mid-B star at z=2.0) and the bright galaxy absolute magnitudes
for H$_0$=65  and q$_0$=0.5 are given in Table 2. These are roughly what we 
expect for L* galaxies with active star-formation. The spatial distribution of
the numerous smaller galaxies suggests they are also associated with 
the group. Their sizes are as small as fractions of an arcsec, corresponding to 
$<$5 Kpc at z=2, and their absolute magnitudes for a young population a z=2 range
from -18 to -20. We note in this respect the group of compact galaxies that
appear to be associated with the z$\sim$2 QSO behind cluster A851 (Dressler et al
1993, Hutchings and Davidge 1997).

\section{Spectra}

   Figure~\ref{fig2} shows the placement of the slits, assuming that the nuclear
slit was centred. The absence of detectable scattered nuclear signal is
consistent with this placement, since the nuclear signal is spread over 
all wavelengths and the resolved light is almost entirely in Ly$\alpha$.
The off-nuclear signals are very weak and are only seen significantly at 
Ly$\alpha$. The faint small galaxy to the SW is excluded from the slit but another
to the W is included. It shows no detectable continuum, which is not
surprising as it is 25 magnitude, but it also shows no line emission in the
wavelength range covered. No other objects are covered by the slits.

   The nuclear spectrum (Figure~\ref{fig5}) 
shows broad line emission from O VI, Ly$\alpha$, N V,
Si IV, C IV, He II and the edge of C III]. There are other weak broad features
(e.g. C III/N IV at 3965A or rest 1307A) and what appear to be several Ly$\alpha$
absorbers. The redshift is measured at 
2.037. The new nuclear spectrum is clearly better than the one
(Wills and Wills 1974) on which the published redshift of 2.039 is based.

   The off-nuclear spectrum was cleaned carefully of hot pixels and cosmic rays
and smoothed with different functions to detect significant signal. No continuum
source is seen in either slit position, including the position of the bright knot
in the radio jet. Figure~\ref{fig6} shows plots of the spectra from different
positions in the images.
 There is faint emission near the position of Ly$\alpha$ extending
several arcsec to the W, and somewhat less to the E of the QSO, in both slit 
positions. The 2 arcsec wide slit means that we have essentially slitless 
spectra of the emission-line
material: the spectral image is determined by the size, position, and velocity
of the resolved regions within the slit. Thus, the undispersed image is essential
in interpreting the spectral image (see e.g. Hutchings et al 1998).

   The slit width corresponds to $\sim$100A or a velocity of 7000 km/s, so the
spectrum is dominated by the spatial structure of the knots, seen in
Figure~\ref{fig2}.
There is a knot of brighter emission at the position along slit A of the
bright radio/optical knot. This Ly$\alpha$ emission is broadened and thus may
indicate a velocity gradient within the knot; however, the signal is weak and 
occupies only a few pixels. No extended signal is detected at C IV.

   In the spectral image from slit A (centred on the nucleus), there are two 
bands of emission, seen on either side of the nuclear peak wavelength
(Figure~\ref{fig6}). Both bands run E-W through the nucleus but the band
at shorter wavelength extends further to the W and less far to the E: it is also
narrower. The slit displaced to the N (slit B) also shows the narrower band, this 
time displaced to longer wavelength. The two slits overlap by 0.5 arcsec just where
there is a region of bright flux seen in the image, so that the apparent wavelength
shift is due to its placement at the left and right side of the two slit positions.
Slit B also shows emission some distance (7 - 8 arcsec) to the E, which does not
correspond to any galaxy or continuum feature in the image. This is shown in 
Figure~\ref{fig6}.
Thus there seems to be gas at the QSO redshift throughout its general vicinity.

   Because of the wide slit, velocities with respect to the nucleus cannot be
estimated accurately. However, using the undispersed image as a template 
(Figure~\ref{fig2}),
we do measure velocities consistently within 100 km/s for the clouds in the
slit overlap region. Thus we have some confidence in the values. They indicate
that clouds to the S of the nucleus have velocities of approach of order 1000 km/s,
while those to the N are receding at velocities of 100 km/s near the nucleus
and up to 1000 km/s to the W.  

If the emission from the brightest knot is correctly identified by its displacement
along the slit, then it, exceptionally,  has high approach velocity, from 
-3200 to -3700 km/s. We see no signal from another other knot
within the radio structure, which lies in slit B, although it is as bright as the
other features in the direct image. All this suggests that Ly$\alpha$
is suppressed in or near the radio jet, and such material has different velocity,
apparently of approach. The Ly$\alpha$ gas not associated with the jet has a
systematic velocity field approaching to the S and receding to the N. 
Finally, we infer that the jet-associated material has a continuum spectrum,
which may be synchrotron or hot stars.
 
   We may make a rough estimate of the flux budget for the host galaxy.
The ratio between nuclear continuum (excluding the strong Ly$\alpha$ and C IV emission
lines) to total image signal, scaled to the same exposure, is 3. That is, some 2/3 
of the image signal comes from the wavelength region from $\sim$5000A to the limit
of the CCD sensitivity near 1 micron. There are no strong emission lines in this
(redshifted) region of the spectrum. The total signal from Ly$\alpha$ extended
line emission is about half the resolved image flux, scaled by this factor 3
(which assumes the SED of the host is the same as the QSO nucleus).  Thus,
it appears that about half the resolved flux in our spectral region is in
the form of continuum radiation which is too faint to detect. The signal level
from this flux spread over several arcsec would be below the detection limits of
the data. Summing 100 rows far from the QSO and comparing them with 100 rows next
to the QSO shows the region near the QSO to have slightly bluer SED - i.e. more 
far-UV continuum. The difference is significant only at the 1-2$\sigma$ level.

   The nuclear spectrum from 3000A to 5700A has total flux of 5.4 x 10$^{-13}$
erg/sec. The measured Ly$\alpha$ flux is $\sim$4 x 10$^{-15}$ erg/sec and from
the image we estimate that about an equal amount lies outside the slits we used,
so that the total Ly$\alpha$ flux is of order 10$^{-14}$ erg/sec. This agrees
reasonably with the value 6.3 x 10$^{-15}$ estimated by Heckman et al (1991).

\section{Conclusion}

   The data presented reveals complex structure of material whose brightest features
correspond with radio knots in a curved compact structure around the QSO. 
We find Ly$\alpha$ cloud velocities up to 1000 km/s near the nucleus, and possibly
several times that within the radio jet. The Ly$\alpha$ flux is weaker relative
to the total image signal in the knots associated with the radio structure,
indicating a higher ionisation in these regions. This is consistent with the
findings of Axon et al (1998) in NGC 1068, and the high ionisation of the regions
of bent radio structure in PKS 2152-699 reported by Fosbury et al (1998).

The QSO
appears to lie at the edge of a dense group of young galaxies with significant
Ly$\alpha$ emitting gas around it. This suggests that the radio/optical structure
is a tail caused by the QSO's motion through the group. We infer that much of the
structure is UV continuum so that the bright knots may be regions of star-formation
excited by the radio jet through the surrounding gas. 

   This QSO and its environment appear to be of considerable interest in offering
a snapshot of early formation of galaxies in a dense group, and the activation of
a QSO. Further information on the dynamics and stellar populations present,
require slit spectroscopy of the galaxies and region with a large aperture
telescope.

  I am grateful to the following for their help and advice on processing the
STIS data: Keith Feggans, Bob Hill, Jon Gardner, and Don Lindler. I also thank
Colin Lonsdale, Daniel Durand, and Sharon Hanna for providing data for some of
the diagrams.

\newpage

\centerline{References}

Axon D.J., Marconi A., Capetti A., Macchetto F.D., Schreier E., Robinson A.,
1998, ApJ, (letters preprint)

Barthel P.D., Miley G.K., Schilizzi R.T., Lonsdale C.J., 1988, AAS, 73, 515

Bremer M.N., Crawford C.S., Fabian A.C., Johnstone R.M., 1992, MNRAS, 254, 614

Dressler A., Oemler A., Gunn J.E., Butcher H., 1993, ApJ, 404, L45

Fosbury R.A.E., Morganti R., Wilson W., Ekers R.D., di Serego Alighieri S., 
Tadhunter C.N., 1998, MNRAS (preprint astro-ph/9801249)

Heckman T.M., Lehnert M.D., van Breugel W., Miley G.K., 1991, ApJ, 370, 78

Hewitt A., and Burbidge G.R., ApJS, 87, 451

Hutchings J.B., and Davidge T.J., 1997, PASP, 109, 667

Hutchings J.B., et al 1998, ApJ, 492, L115

Lehnert M.D., Heckman T.M., Chambers K.C., Miley G.K., 1992, ApJ, 393, 68

Lonsdale C.J., Barthel P.D., Miley G.K., 1993, ApJS, 87, 63

Wills D., and Wills B.J., 1974, ApJ, 190, 271

\newpage

\centerline{Captions to figures}

\figcaption[hutchings.fig1.ps]{STIS CCD image of QSO 1345+584 and surroundings. 
The QSO and its brightest 6
companions are labelled. N is up and E to the left, and the field shown is 
40 x 30 arcsec. The bright `arm' to the E of the QSO is part of the PSF. The field
outside the section shown is empty of galaxies on this display. \label{fig1}}

\figcaption[hutchings.fig2.ps]{Detail of QSO image before and after PSF subtraction.
The area shown is 7.5 arcsec on a side. The subtracted image is
smoothed to reduce noise. The two slit positions are shown: they are 2 arcsec wide
and offset by 1.5 arcsec, and thus overlap by 0.5 arcsec. The X marks the nucleus
position. \label{fig2}}

\figcaption[hutchings.fig3.ps]{Contour plots of STIS image matched with 
ground-based data. 
Upper: smoothed STIS CCD image on left and optical PSF-subtracted image from
Heckman et al (1991). Lower: PSF-subtracted STIS image and 15 GHz image from 
Lonsdale et al (1993). The QSO nucleus is marked with X and the radio
knots correspond with optical knots. See text for discussion. \label{fig3}}

\figcaption[hutchings.fig4.ps]{Azimuthally averaged luminosity profile of 
1345+584 and star image at same
point on the detector. Typical error bars are shown at different radii, but the 
resolved flux is not azimuthally symmetric. \label{fig4}}

\figcaption[hutchings.fig5.ps]{Spectrum of the QSO nucleus with principal 
emission lines marked. Note the
unreddened continuum and Ly$\alpha$ absorption lines. \label{fig5}}

\figcaption[hutchings.fig6.ps]{Off-nuclear spectra compared with the nucleus, 
which has been scaled by 1/100. The knot spectrum is scaled by 1/3 of the others.
The units are non-fluxed counts. The lowest traces are smoothed sky-subtracted
plots, with approximate linear normalisations sketched in. Note the different
Ly$\alpha$ emission components. Most of the line displacement in wavelength is
due to positions of line emitting regions within the 2" wide slits. \label{fig6}}

\newpage

\begin{deluxetable}{llll}
\tablenum{1}
\tablecaption{STIS observations of QSO 1345+584}
\tablecaption{Position angle -98$^o$, Dec 19 1997}
\tablehead{\colhead{Observation} &\colhead{Mode} &\colhead{Slit} 
&\colhead{Exposure} \\ &&\colhead{(sec)}}
\startdata
Image  &MIRVIS &50CCD   &200\nl
Image &MIRVIS &50CCD &2000\nl
Spectrum on-nucleus &G450L &52x2 &3360\nl
Spectrum 1.5" N of nucleus &G450L &52x2 &3360\nl
\enddata
\end{deluxetable}

\begin{deluxetable}{llll}
\tablenum{2}
\tablecaption{Galaxies in field of QSO 1345+584}
\tablehead{\colhead{Galaxy \#} &\colhead{Comment} &\colhead{Mag} 
&\colhead{Abs mag\tablenotemark{a}}}
\startdata
Q &QSO &18.5 &-26.0\nl
Q &Fuzz &22.4 &-22.1\nl
1 &Edge-on disk &21.7 &-22.8\nl
2 &Knotty + star &22.7 &-21.8\nl
3 &LSB smooth with knot &22.1 &-22.3\nl
4 &Compact knotty &22.7 &-21.8\nl
5 &Face-on knotty spiral &22.0 &-22.5\nl
6 &Nucleated E with knot &21.8 &-22.7\nl
&Typical small galaxy &24.5 &-20\nl
\enddata
\tablenotetext{a}{H$_0$=65, q$_0$=0.5}
\end{deluxetable} 

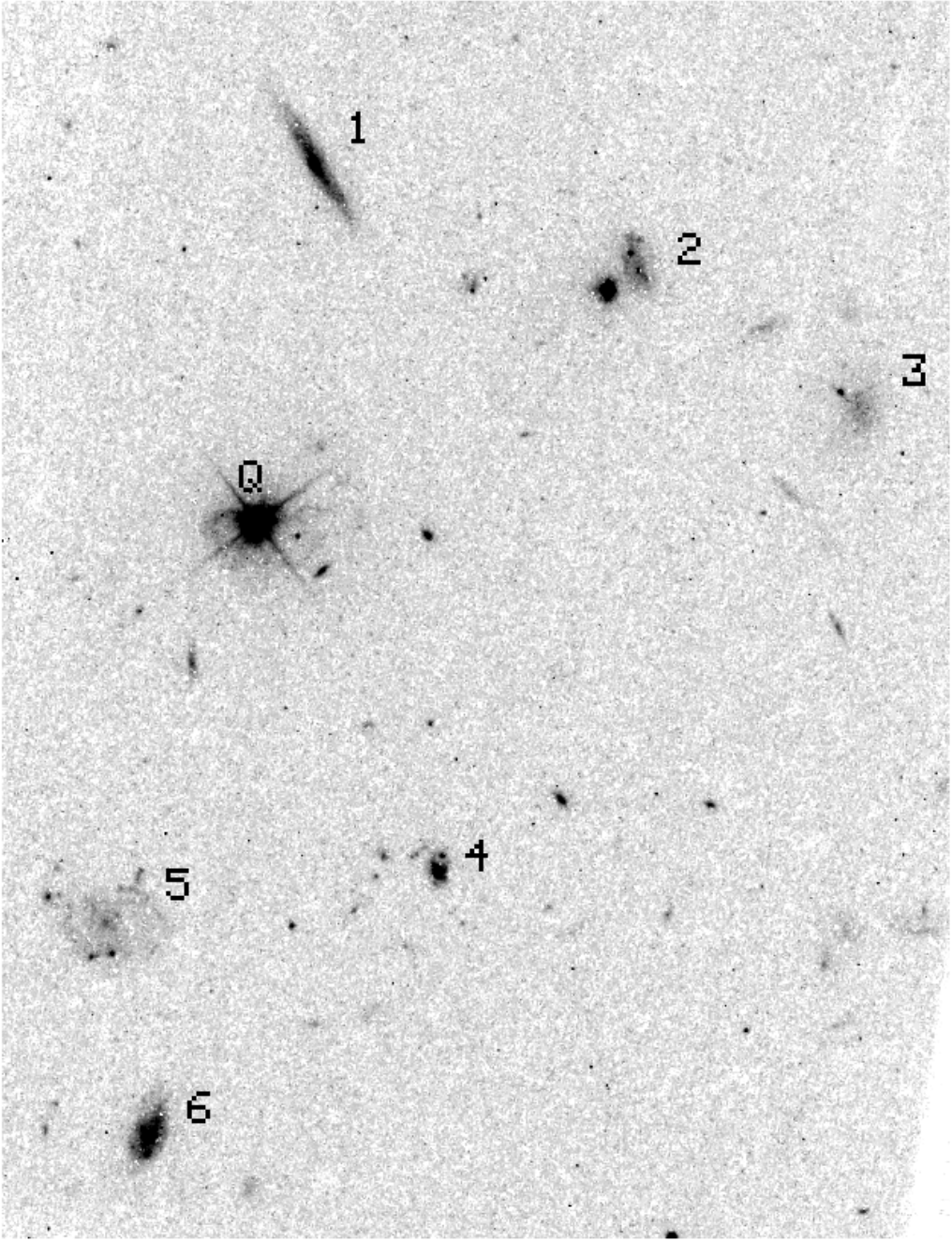
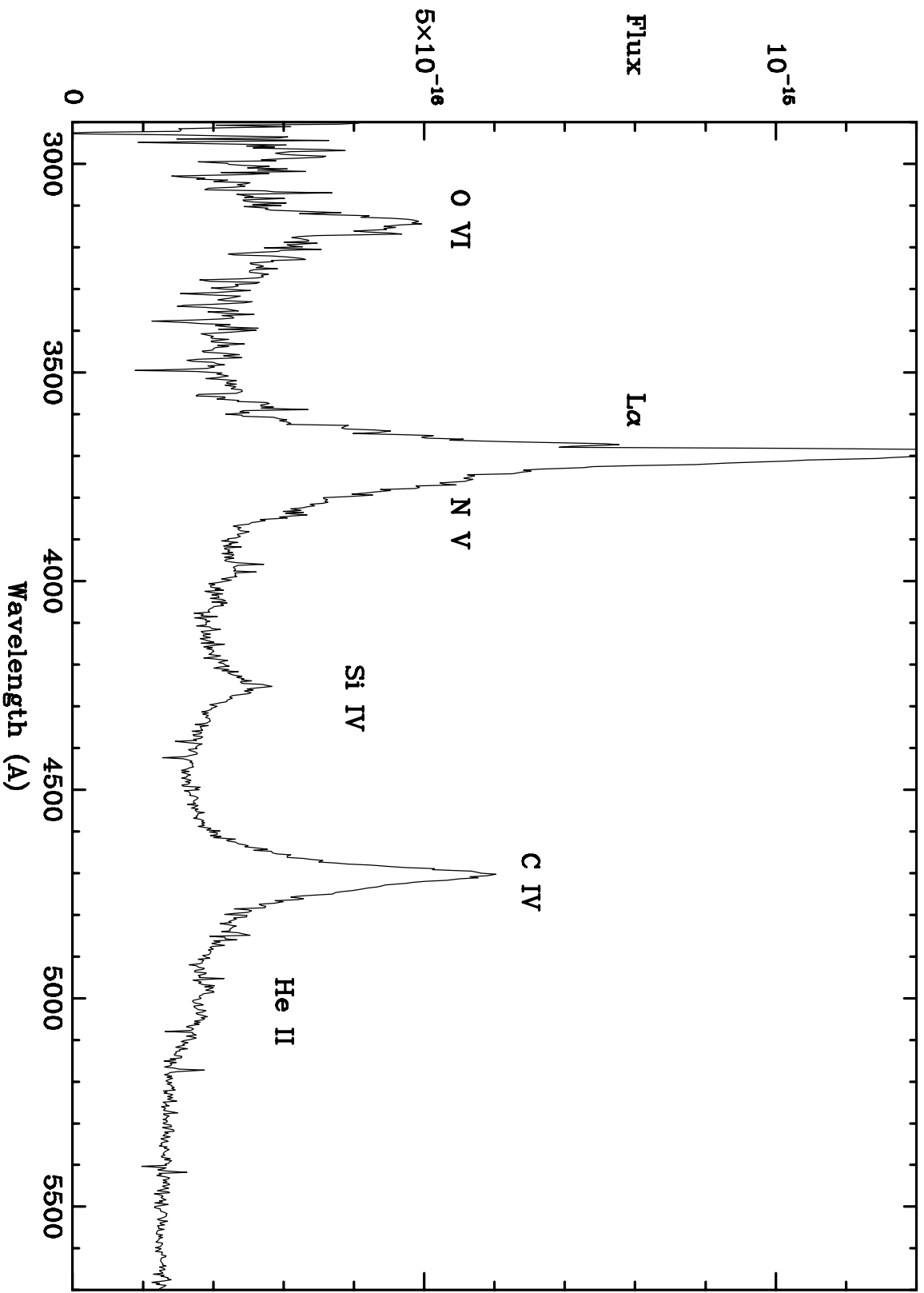
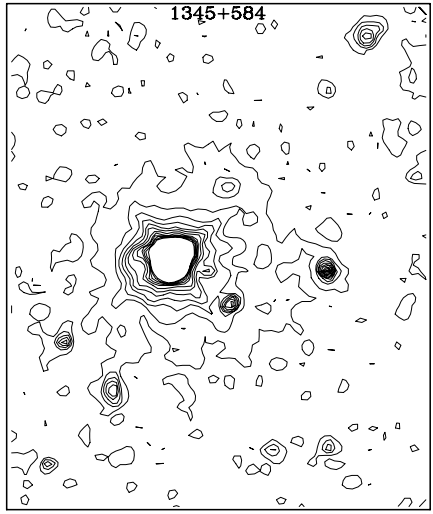
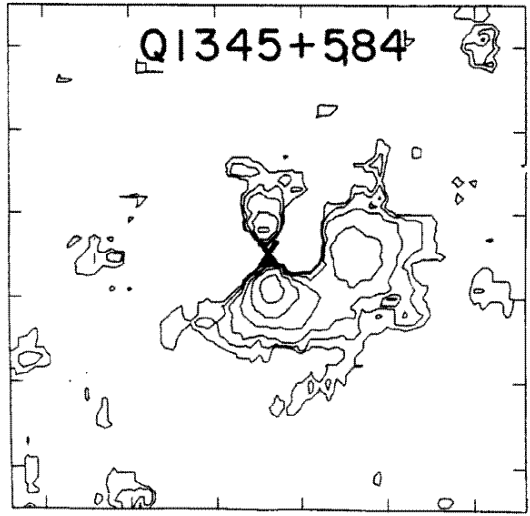
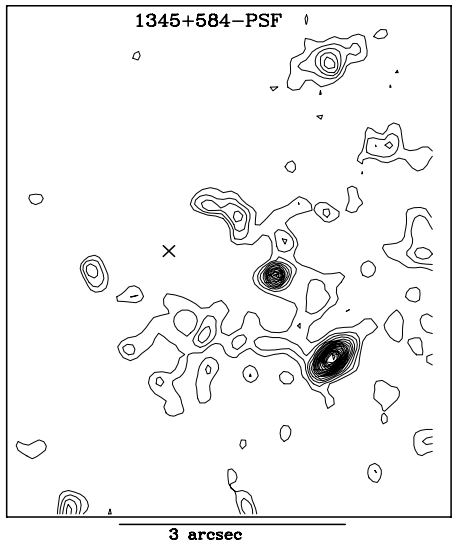
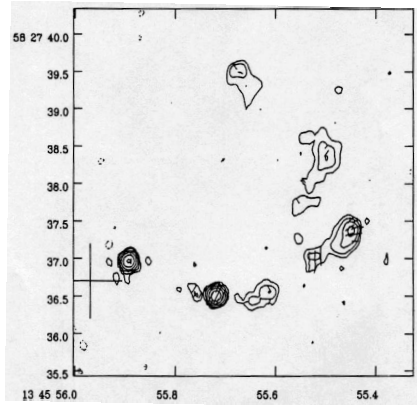
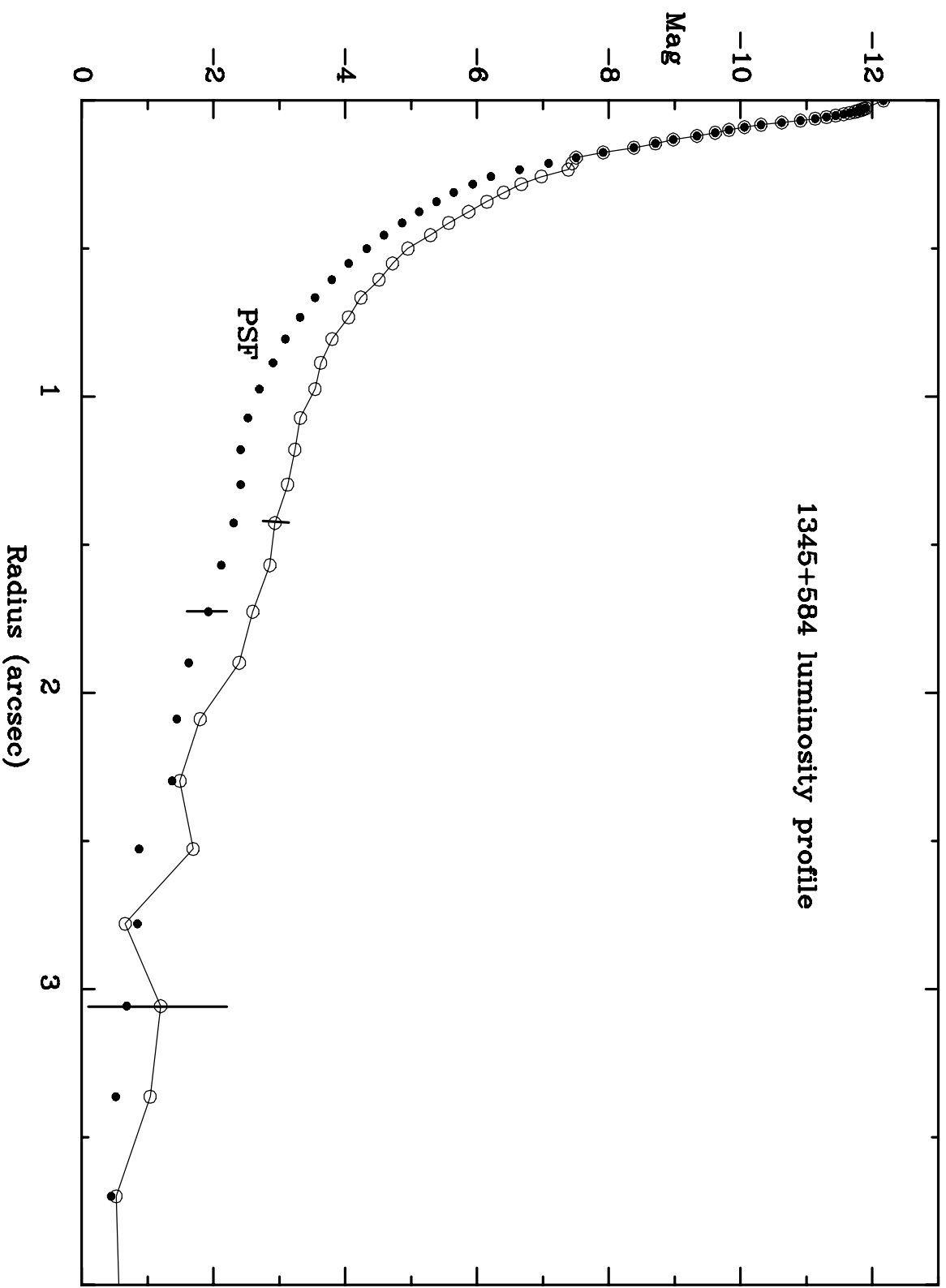
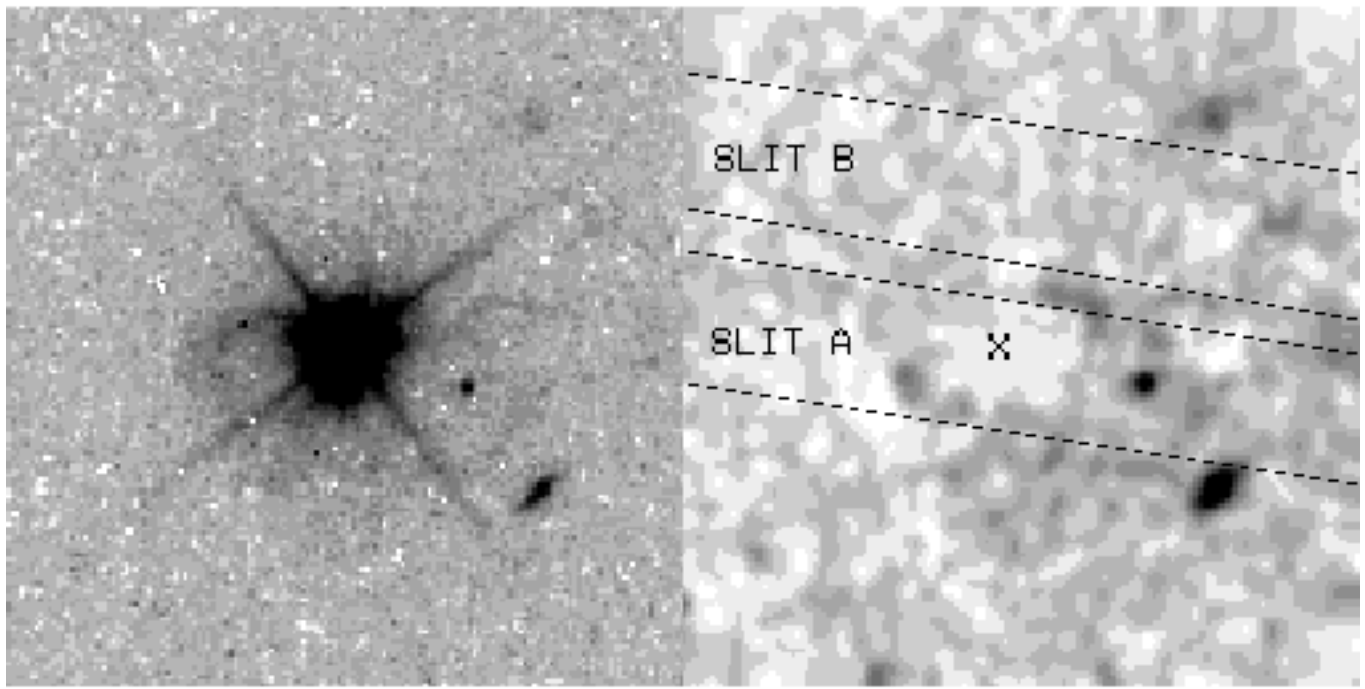
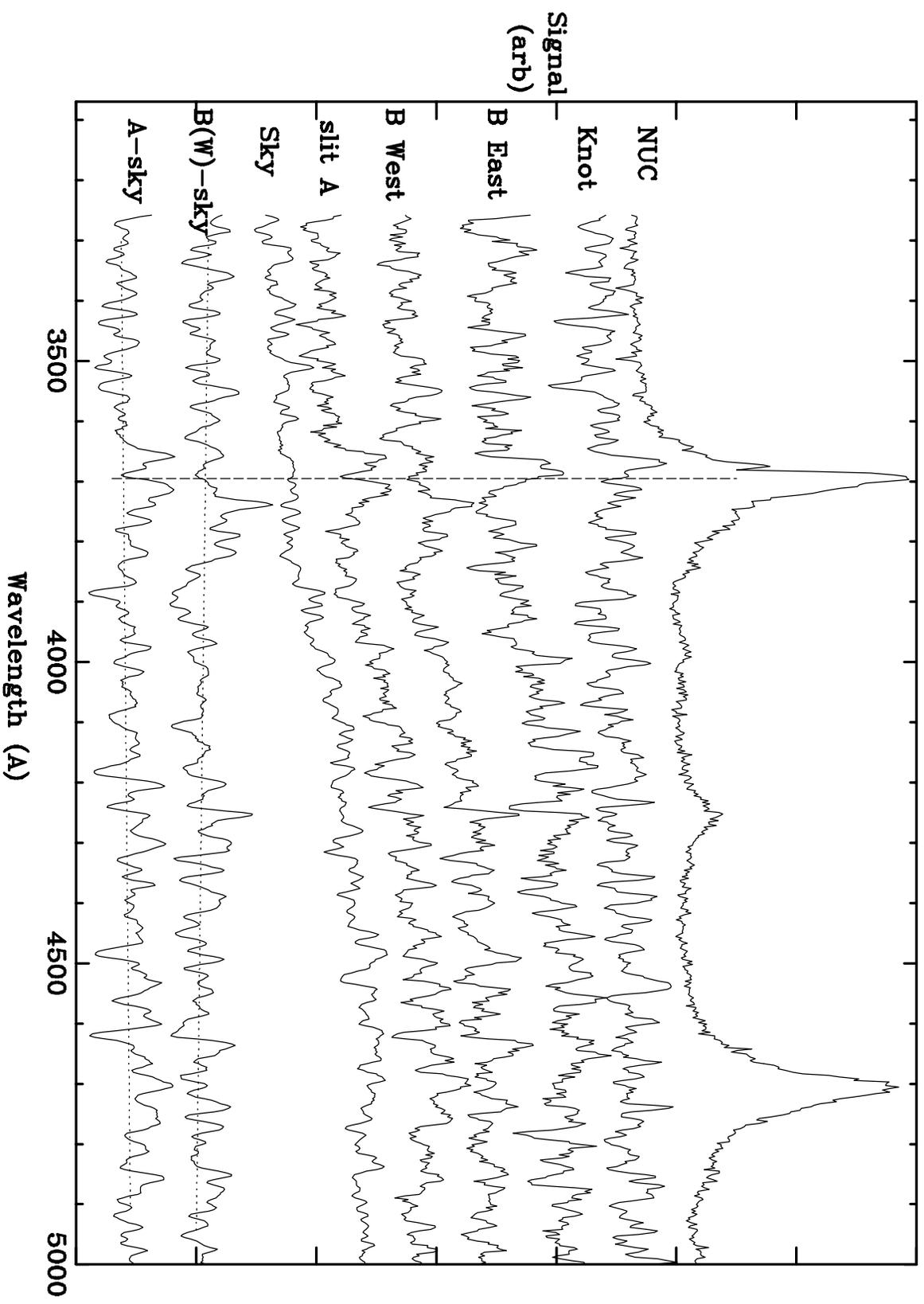
\end{document}